\documentclass[twocolumn]{aastex6}

\newcommand\simlt{\lower.5ex\hbox{$\; \buildrel < \over \sim \;$}}
\newcommand{\fermi}{{\it Fermi}-LAT }
\newcommand{\gray}{$\gamma$-ray }
\newcommand{\grays}{$\gamma$-rays }

\collaborationName{Friends of AASTeX}

\begin{document}


\title{Rapid Gamma-ray variability of NGC 1275}

%% Use \author, \affil, and the \and command to format author and affiliation
%% information.  If done correctly the peer review system will be able to
%% automatically put the author and affiliation information from the manuscript
%% and save the corresponding author the trouble of entering it by hand.
%%
%% The \affil should be used to document primary affiliations and the
%% \altaffil should be used for secondary affiliations, titles, or email.

%% Authors with the same affiliation can be grouped in a single
\author{V. Baghmanyan$^{1}$, S. Gasparyan$^{1,2}$ and N. Sahakyan$^{1,2}$}
\affil{$^{1}$ ICRANet Armenia Marshall Baghramian Avenue 24a, 0019\\
Yerevan, Republic of Armenia\\
$^{2}$ ICRANet Piazza della Repubblica 10, 65122\\
 Pescara, Italy}


\altaffiltext{1}{narek@icra.it}
%t%% AASTeX 6.0 supports the ability to suppress the names and affiliations
%% of some authors and displaying them under a "collaboration" banner to
%% minimize the amount of author information that to be printed.  This
%% should be reserved for articles with an extreme number of authors.
%% The necessary command are \AuthorCallLimit and \collaborationName.
%% An \AuthorCallLimit=2 call prior to the author list will only show
%% the authors in the first two \author calls.  The \collaborationName
%% defines the collaboration identifier.  Commented examples below.

%\AuthorCallLimit=1
%% Will only show Schwarz & Muench since Schwarz and Muench
%% are in the same \author call.
%\collaborationName{Friends of AASTeX}
%% will print "The AAS collaboration" after the shortened author list.
%% Note that all the \altaffil information will still be shown so it
%% has to be manually commented out if you do not want it shown.
%%
%% Note that all of these author will be shown in the published article.
%% This feature is meant to be used prior to acceptance to make the
%% front end of a long author article more manageable.

\begin{abstract}
We report on a detailed analysis of the $\gamma$-ray light curve of NGC 1275 using the Fermi large area telescope data accumulated in 2008-2017.  Major $\gamma$-ray flares were observed in October 2015 and December 2016/January 2017 when the source reached a daily peak flux of $(2.21\pm0.26)\times10^{-6}\:{\rm photon\:cm^{-2}\:s^{-1}}$, achieving a flux of $(3.48\pm0.87)\times10^{-6}\:{\rm photon\:cm^{-2}\:s^{-1}}$ within 3 hours, which corresponds to an apparent isotropic $\gamma$-ray luminosity of $\simeq3.84\times10^{45}\:{\rm erg\:s^{-1}}$. The most rapid flare had e-folding time as short as $1.21\pm0.22$ hours which had never been previously observed for any radio galaxy in $\gamma$-ray band. Also $\gamma$-ray spectral changes were observed during these flares: in the flux versus photon index plane the spectral evolution follows correspondingly a counter clockwise and a clockwise loop inferred from the light curve generated by an adaptive binning method. On December 30, 2016 and January 01, 2017 the X-ray photon index softened ($\Gamma_{\rm X}\simeq 1.75-1.77$) and the flux increased nearly $\sim3$ times as compared with the quiet state. The observed hour-scale variability suggests a very compact emission region ($R_\gamma\leq5.22\times10^{14}\:(\delta/4)\:{\rm cm}$) implying that the observed emission is most likely produced in the subparsec-scale jet if the entire jet width is responsible for the emission. During the active periods the $\gamma$-ray photon index hardened, shifting the peak of the high energy spectral component to $>{\rm GeV}$, making it difficult to explain the observed X-ray and $\gamma$-ray data in the standard one-zone synchrotron self-Compton model.
\end{abstract}


\section{Introduction} \label{sec:intro}
Due to its proximity ($z=0.0176$, $\approx 75.6$ Mpc) and brightness, the radio galaxy NGC 1275 has been a target for observations in almost all energy bands. Core-dominated asymmetrical jets at both kpc \citep{pedlar} and pc scales \citep{asada} have been detected in the radio band with characteristics more similar to those of Fanaroff and Riley type 1 sources \citep{fanarof}. The emission in the X-ray band is mostly dominated by the thermal emission from the cluster, although a nonthermal component in the energy range  0.5-10 keV with a photon index of $\Gamma_{\rm X}\simeq1.65$ has been observed \citep{churazov,fabian}. High Energy (HE; $>100\:{\rm MeV}$) \grays from NGC 1275 had already been detected by Fermi Large Area Telescope (\fermi) using the data obtained during the first 4 months of observations \citep{abdo2009}. Then, using the data accumulated for longer periods \gray flux and photon index variation on month timescales were detected \citep{kataoka2010}. However, the \gray emission is variable also in shorter (a few days’) timescales \citep{brown}. Very High Energy (VHE; $>100\:{\rm GeV}$) \gray emission with a steep spectral index of $4.1\pm0.7$ was detected by MAGIC, using the data accumulated between August 2010 and February 2011 \citep{aleksic2012}. No hints of variability above $100$ GeV were seen on month time scales.\\
Even if the observed \gray variability allowed to exclude Perseus cluster as the main source of \gray emission, the exact mechanisms responsible for the broadband emission from NGC 1275 are still unclear. The multiwavelength Spectral Energy Distribution (SED) hints at a double-peaked SED with the peaks around $10^{14}\:{\rm Hz}$ and $(10^{23}-10^{24})\:{\rm Hz}$ \citep{aleksic2014}. Within a ''classical'' misaligned BL Lac scenario, a one-zone synchrotron/Synchrotron Self Compton (SSC) interpretation of the SED can well explain the HE peak constrained by \fermi and MAGIC data but has difficulties explaining the low energy data. It requires that the jet should be more aligned than it is estimated, e.g., $30^\circ-55^\circ$ \citep{walker}. Therefore additional assumptions on the jet properties and/or more complex scenarios for inverse-Compton scattering should be made.\\
In the HE \gray band frequently flaring activities are known for NGC 1275 \citep{baghmanyan}. A substantial increase of the \gray flux in the HE and VHE \gray bands was detected in October 2015 and January 2017 \citep{pivato, mirzoyan, mukherjee, lucarelli}. In October 25, 2015 \fermi detected a bright flare with a daily peak flux of $(1.6\pm0.2)\times10^{-6}\:{\rm photon\:cm^{-2}\:s^{-1}}$ \citep{pivato}. Then, in the night between 31 December 2016 and 01 January 2017 a major flare was detected in the VHE \gray band when the flux was 60 times higher than the mean flux \citep{mirzoyan}. Also the flux $>100$ MeV was about 12 times higher than the most significant flux observed with AGILE \citep{lucarelli}. Besides, Swift observations during this major \gray active period provided data in the UV and X-ray bands and so giving a unique chance to investigate the flaring activity of NGC 1275 in the multiwavelength context.\\
The goal of this paper is to have a new look on the \gray emission from NGC 1275 in the last $\sim8.7$ years in general and during the major flaring periods in particular. The larger data set allows to investigate the \gray flux evolution in time with improved statistics in shorter time scales, while a detailed analysis of recently observed exceptional flares will allow to have an insight into the particle acceleration and emission processes.\\
This paper is organized as follows. The \fermi data reduction and temporal analyses are presented in Section \ref{sec2}. The spectral analyses of \fermi and Swift data are described in Section \ref{sec3}. We present our results and discussion in Section \ref{sec4}. Summary is given in Section \ref{sec5}.
\begin{figure*}
  \centering
    \includegraphics[width= 0.96 \textwidth]{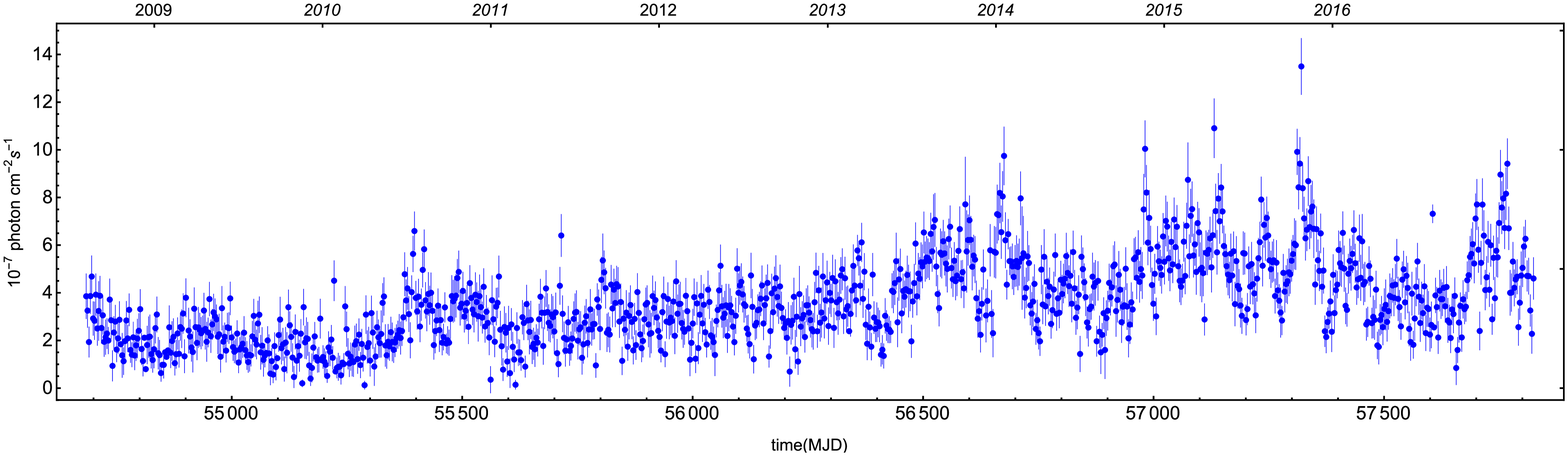}\\
     \includegraphics[width= 0.48 \textwidth]{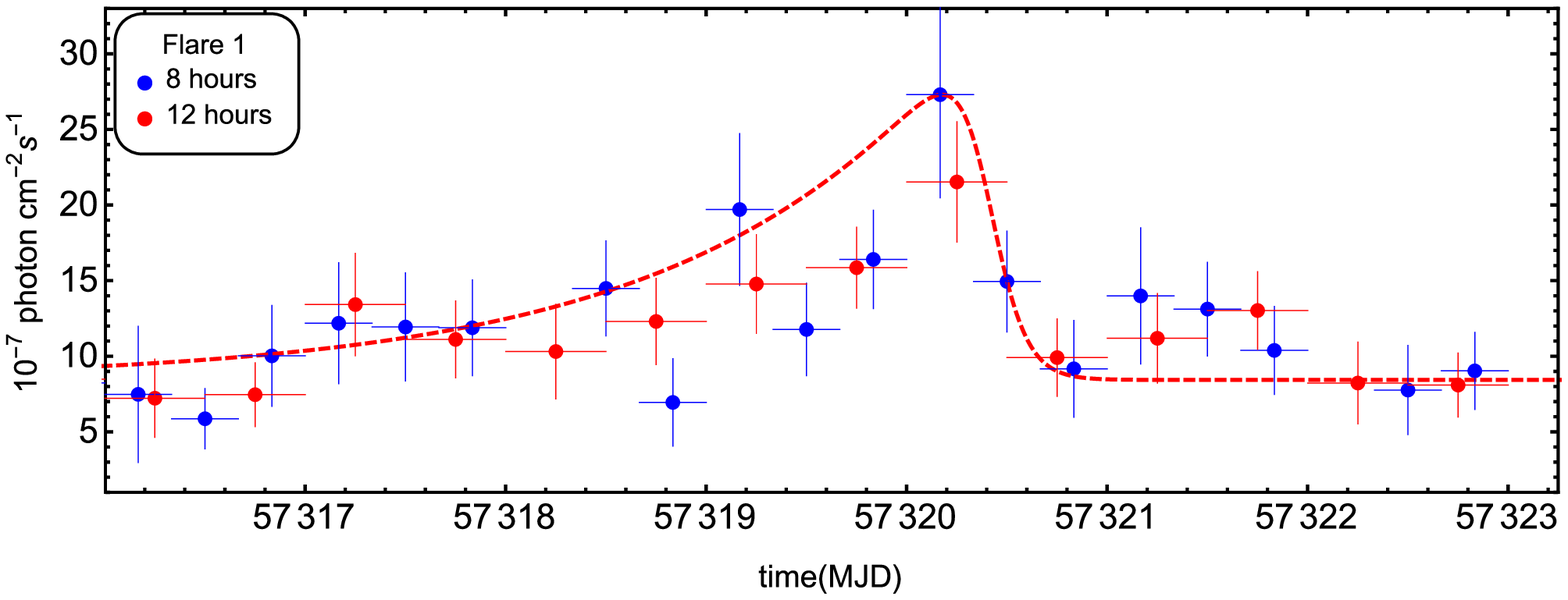}
     \includegraphics[width= 0.48 \textwidth]{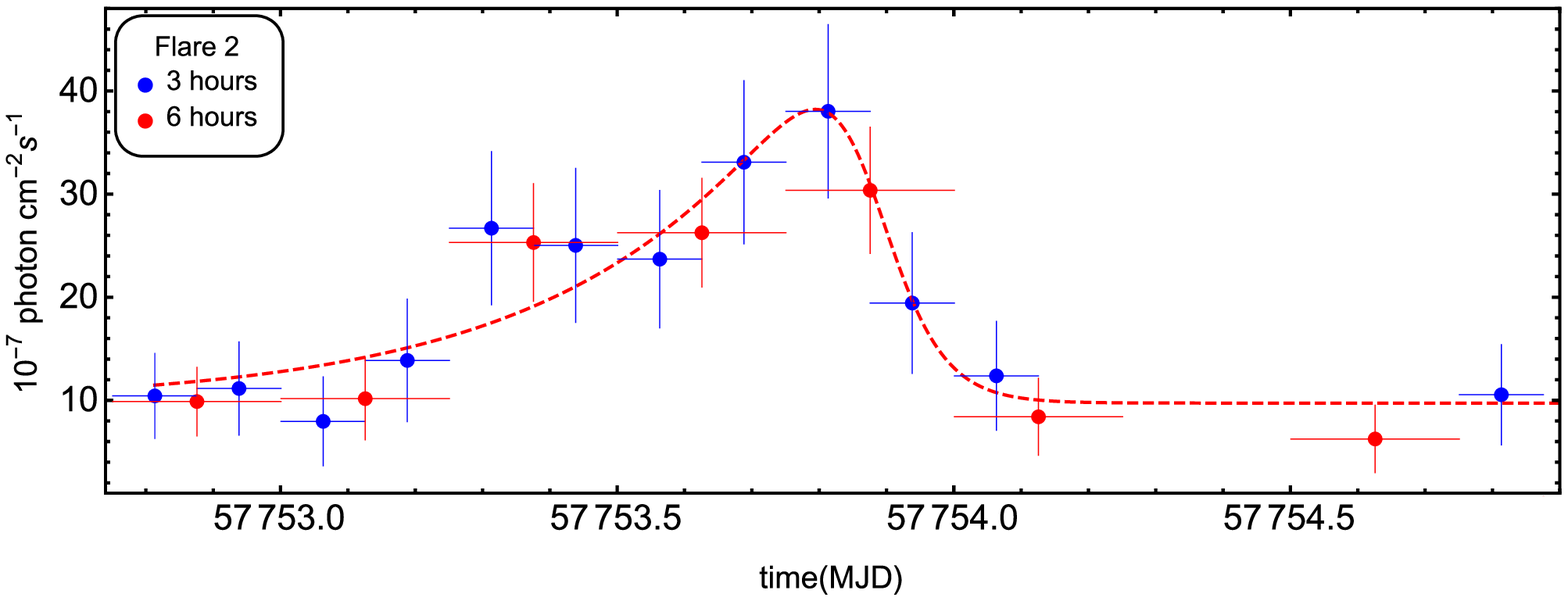}\\
     \includegraphics[width= 0.478 \textwidth]{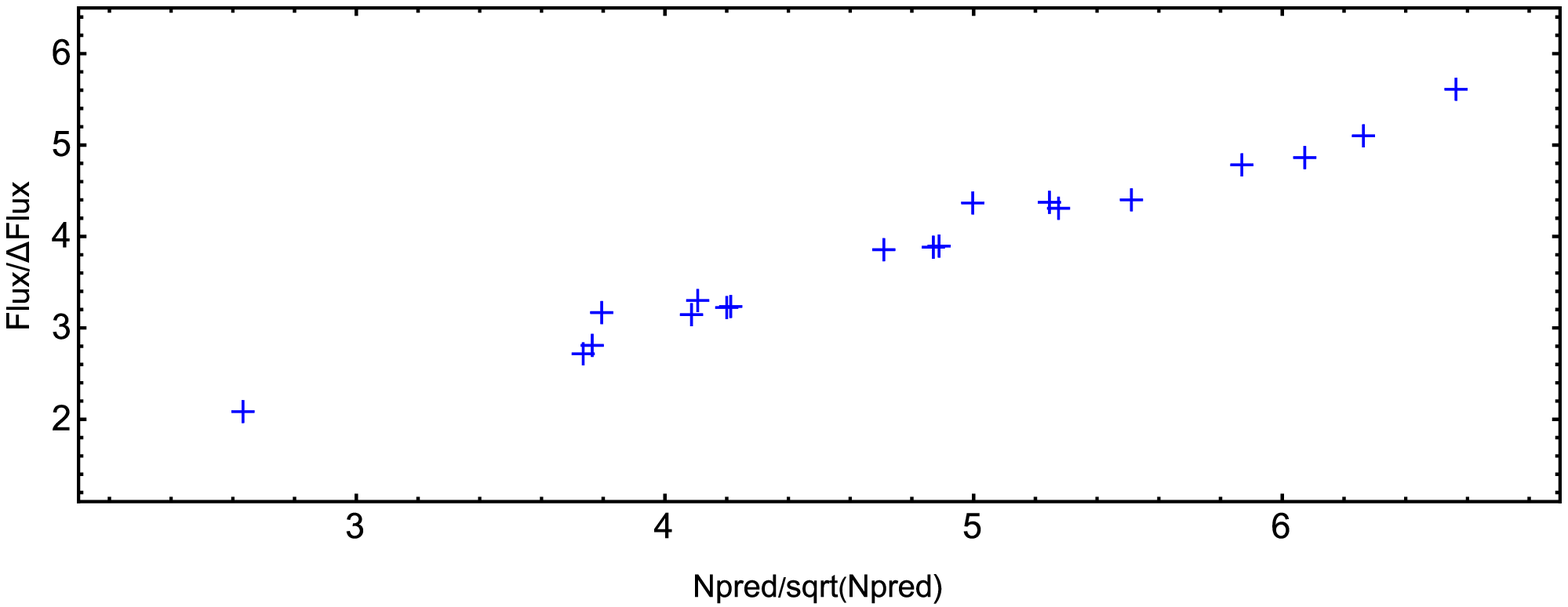}
     \includegraphics[width= 0.478 \textwidth]{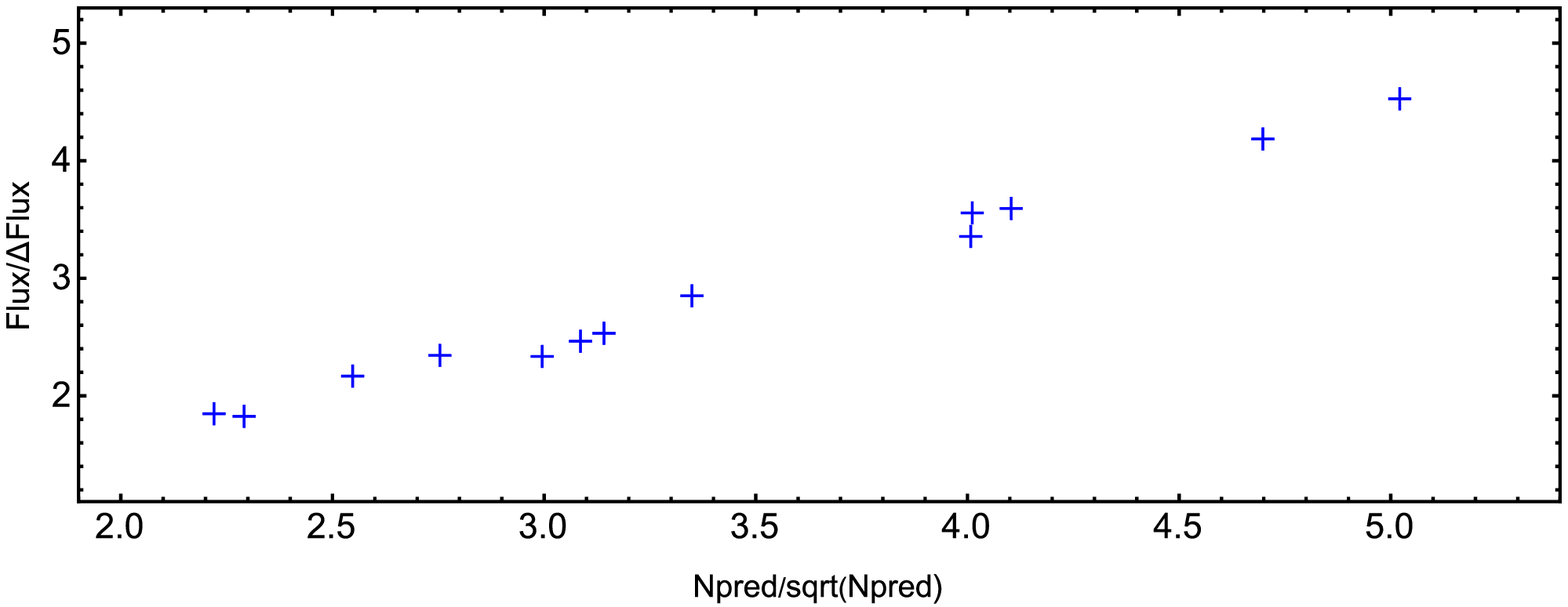}
    \caption{{\it Top panel:} The light curve of $0.1<\:E_\gamma\:<500$ GeV \grays from NGC 1275 from  August 4, 2008 to  March 5, 2017, with 3-day (blue) binning. {\it Middle panels:} Sub intervals covering F1 (left) and F2 (right). F1 is shown with 8-hour (blue) and 12-hour (red) time intervals and F2 with 3-hour (blue) and 6-hour (red) bins. The red dashed lines show the fit of F1 and F2 with Eq. \ref{func}. {\it Lower panels:} The plot of Npred/$\sqrt{\rm Npred}$ vs Flux/$\Delta {\rm Flux}$ for 8-hour (left) and 3-hour (right) bins.}%
%    %\vspace{-1.6em}
    \label{fg2}
\end{figure*}
\begin{table*}[t!]
\scriptsize
 \begin{center}
 \caption{Parameters of spectral analysis}\label{tab:results}
 \resizebox{0.9 \textwidth}{!}{
 \begin{tabular}{c c c c c c}
 \hline
 \hline
  \multicolumn{6}{c}{\fermi} \\
 \hline
  Period & Date & Flux\tablenotemark{a} & Photon Index \tablenotemark{b} & Test Statistic & Highest photon energy\tablenotemark{c}  \\
  \hline
  57442.32-57444.45 & 2016/02 (24$-$26) & $4.18\pm0.85$ & $1.93\pm0.14$ & 123 & 10.39 \\
  57752.75-57753.25 & 2016/12 (30$-$31) & $8.56\pm2.30$ & $1.79\pm0.17$ & 106 & 34.77\\
  57753.81 & 2016/12 31 & $34.82\pm8.67$ & $1.93\pm0.19$ & 102 & 5.84 \\
  57754.00-57755.75 & 2017/01 (01$-$02)  & $6.27\pm1.20$ & $1.67\pm0.11$ & 178 & 4.18 \\
 \hline
 \multicolumn{6}{c}{Swift-XRT} \\
 \hline
  Obsid & Date & Exp. time & Photon Index \tablenotemark{d} & Unabsorbed Flux \tablenotemark{e} & $\chi^2_{\rm red}$ (d.o.f.) \\
  \hline
  34380005 & 2016-02-25  & 2750 & 1.52 $\pm$ 0.08 & 3.10 $\pm$ 0.19 & 1.04 (75) \\
  87312001 & 2016-12-30  &  939 & 1.75 $\pm$ 0.12 & 8.64 $\pm$ 0.76 & 0.74 (24) \\
  87311001 & 2017-01-01  & 619 & 1.77 $\pm$ 0.17 & 10.57 $\pm$ 1.26 & 1.15 (15) \\
  31770011 & 2017-01-01  & 984 & 1.77 $\pm$ 0.08 & 10.29 $\pm$ 0.55 & 0.95 (173) \\
  \hline
\multicolumn{6}{l}{%
  \begin{minipage}{0.85 \textwidth}
 \tablenotetext{} {\bf Notes:}
 \tablenotetext{a}{Integrated \gray flux in the $0.1-100$ GeV energy range in units of $10^{-7}\:{\rm photon\:cm^{-2}\:s^{-1}}$.}
 \tablenotetext{b}{\gray photon index from likelihood analysis.}
 \tablenotetext{c}{Photon energy in GeV.}
 \tablenotetext{d}{Photon index from X-ray data analysis.}
 \tablenotetext{e}{X-ray flux in the energy range 0.3--10 keV in units of $\times$10$^{-11}$ erg cm$^{-2}$ s$^{-1}$ (corrected for the Galactic absorption).}
\end{minipage}%
}\\
 \end{tabular}}
\end{center}

\end{table*}
\section{FERMI LAT OBSERVATIONS AND DATA ANALYSIS}\label{sec2}
The Large Area Telescope on board the Fermi satellite is a pair-conversion telescope sensitive to \grays in the energy range of 20 MeV - 500 GeV \citep{atwood2009}. We have used the publicly available data accumulated during the last $\sim8.7$~years of \fermi operation (from August 4, 2008 to March 15, 2017). The data were analyzed with the standard Fermi Science Tools v10r0p5 software package released on May 18, 2015. The most recent reprocessed PASS 8 events and spacecraft data were used with {\it P8R2\_SOURCE\_V6} instrument response function. Only the events with a higher probability of being photons (evclass=128, evtype=3) in the energy range of 100 MeV - 500 GeV were analyzed. In the analysis we selected different radii ($9^\circ$, $10^\circ$, $12^\circ$ and $15^\circ$) of the Region of Interest (ROI) to ensure that the selected ROI is an accurate representation of the observation. This yielded essentially the same results within statistical uncertainties, so a radius of  $12^\circ$ was used and the photons from a $16.9^{\circ}\times16.9^{\circ}$ square region centered at the location of NGC 1275, (RA,dec)= (49.96, 41.51), were downloaded. The recommended quality cuts, (DATA\_QUAL==1)$\&\&$(LAT\_CONFIG==1) and a zenith angle cut at $90^\circ$ to eliminate the Earth limb events were applied with {\it gtselect} and {\it gtmktime} tools. We binned photons with {\it gtbin} tool with an Aitoff projection into pixels of $0.1^\circ\times0.1^\circ$ and into 37 equal logarithmically-spaced energy bins. Then with the help of {\it gtlike} tool, a standard binned maximum likelihood analysis is performed. The fitting model includes diffuse emission components and \gray sources within ROI. The model file is created based on the \fermi third source catalog (3FGL) \citep{3fgl} and the Galactic background component is modeled using the Fermi LAT standard diffuse background model {\it gll\_iem\_v06} and {\it iso\_P8R2\_SOURCE\_V6\_v06} for the isotropic \gray background. The normalization of background models as well as fluxes and spectral indices of sources within $12^{\circ}$ are left as free parameters. As in 3 FGL, the \gray spectrum of NGC 1275 was modeled using a log-parabolic spectrum.\\
Using of the data accumulated for an almost 2 times longer period than in 3 FGL, can result in new \gray sources in the ROI which are not properly accounted for in the model file. In order to probe for additional sources, a Test Statistics map of the ROI (TS defined as TS $= 2(log {\rm L}-log {\rm L_0})$, where ${\rm L}$ and ${\rm L_0}$ are the likelihoods when the source is included or is not) is created with {\it gttsmap} tool using the best-fit model of 0.1-500 GeV events. To identify the coordinates of  the excess hot spots with TS $>$ 25 ($5\:\sigma$) we used the {\it find\_source} iterative source-finding algorithm implemented in {\it Fermipy} \footnote{http://fermipy.readthedocs.io/en/latest/}. In the TS map it identifies the peaks with ${\rm TS} >25$ and adds a source at each peak starting from the highest TS peak. The sources position is obtained by fitting a 2D parabola to the log-likelihood surface around the maximum. Alternatively, the sources position was calculated by hand using the pixels surrounding the highest TS (similar to the method used in \citet{Macias}). Both methods resulted in similar values. For each given point we sequentially added a new point source with a conventional spectral definition (power-law) and performed binned likelihood analysis with {\it gtlike}. For the further analysis we used the model file with the new additional point-like sources to have better representation of the data.
\subsection{Temporal variability}
\begin{figure*}
  \centering
    \includegraphics[width= 0.96 \textwidth]{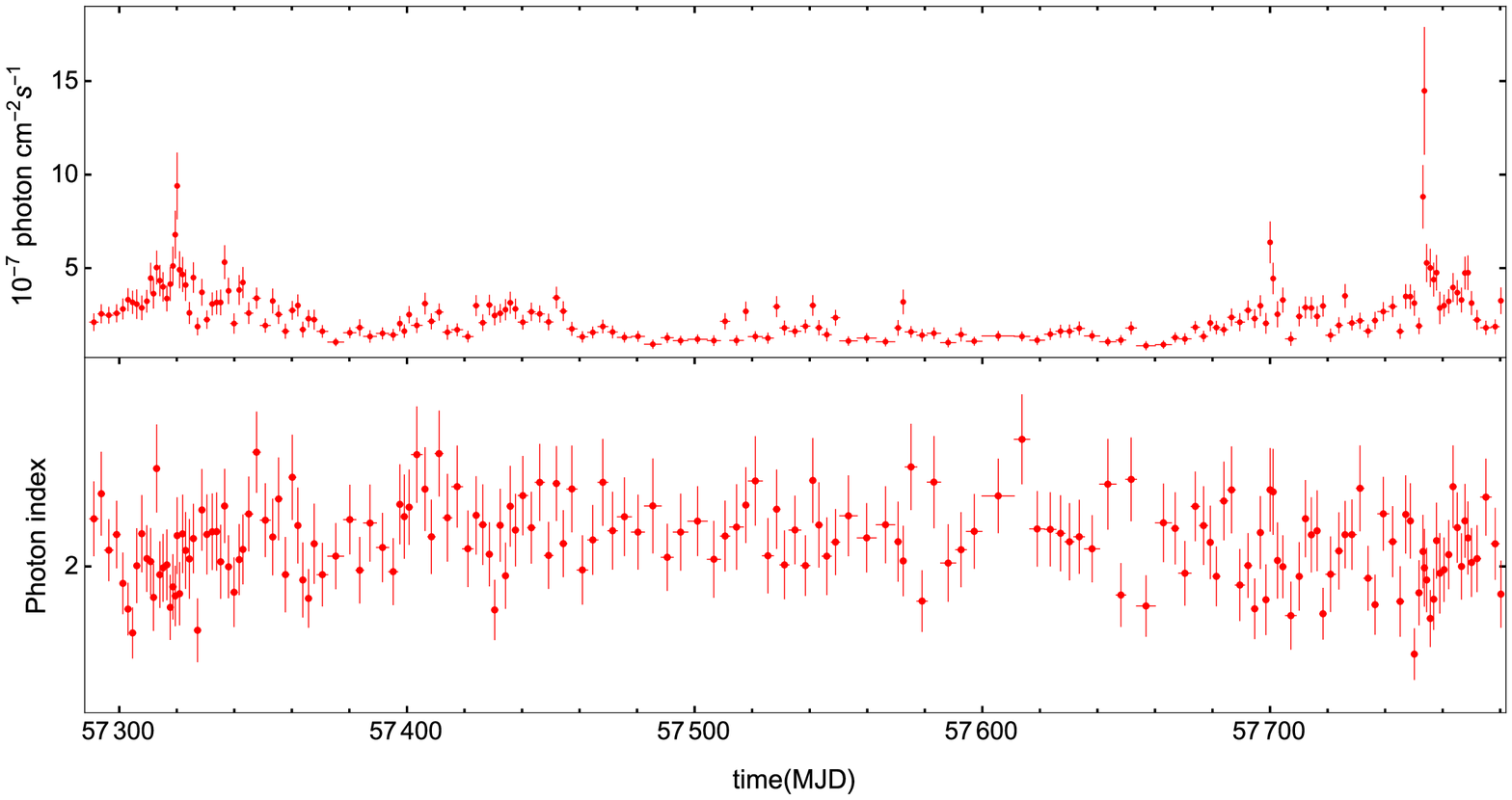}
      \includegraphics[width= 0.48 \textwidth]{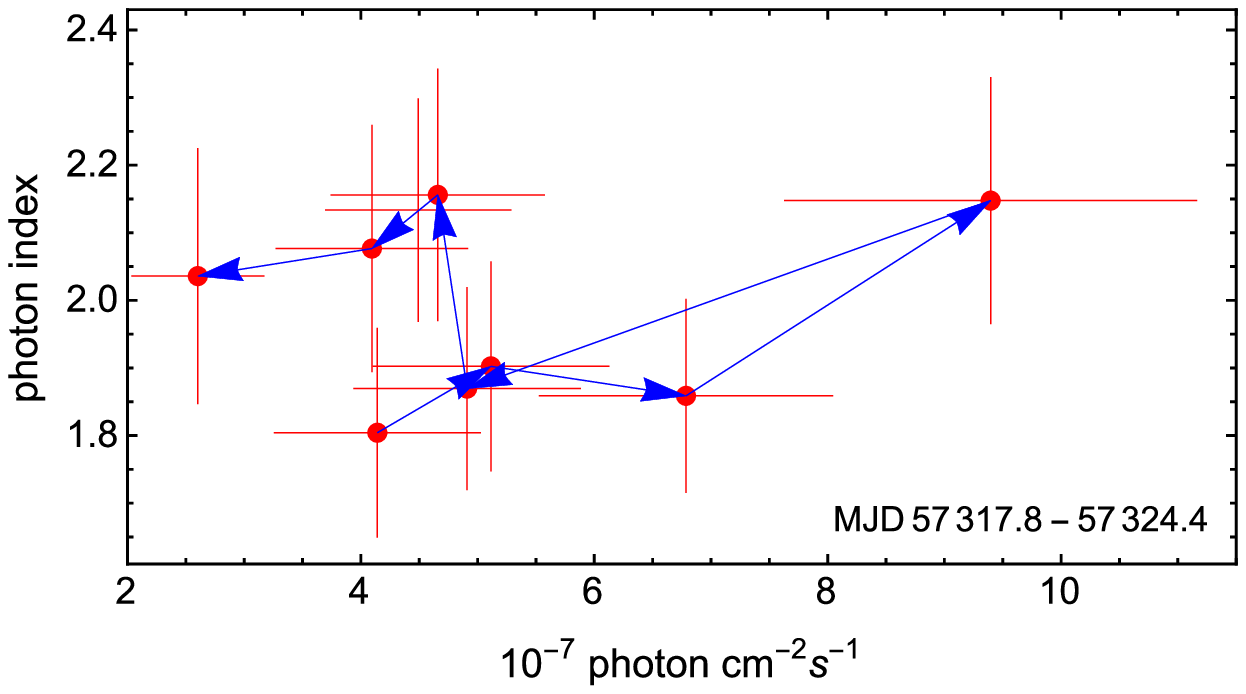}
      \includegraphics[width= 0.48 \textwidth]{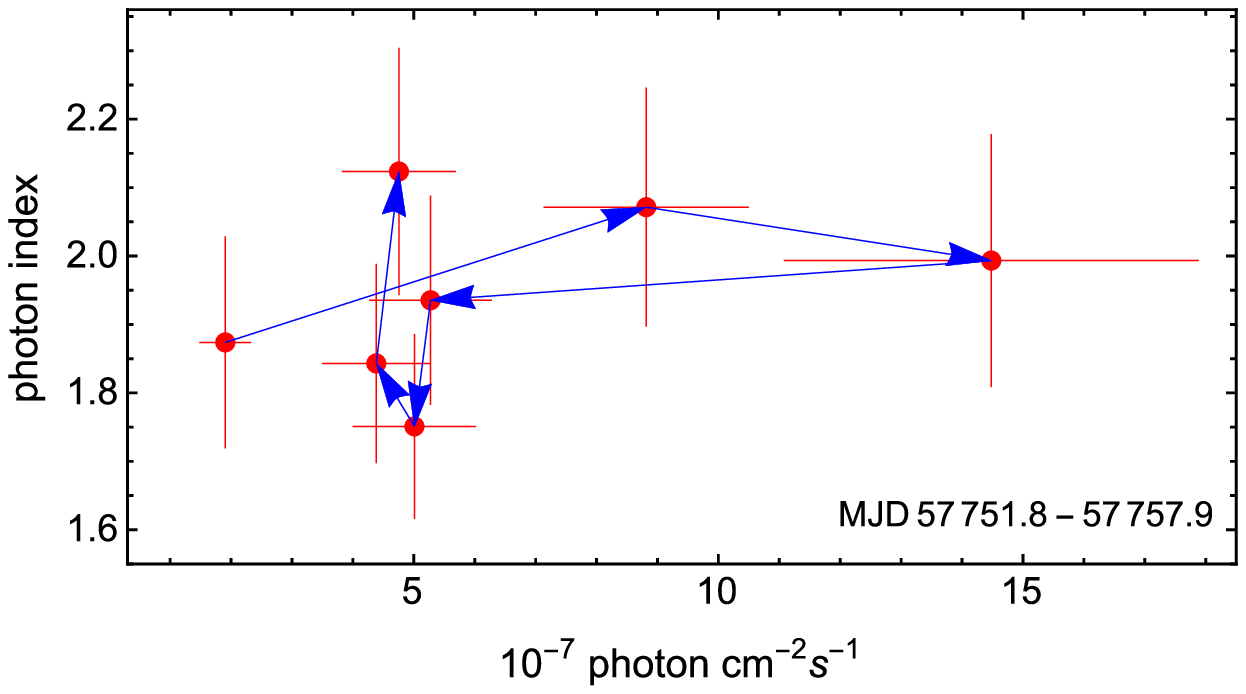}
    \caption{{\it Top panels:} The upper panel shows the period of major flares with the constant uncertainty (15\%) light curve above $208.6$ MeV obtained with the adaptive binning method. The lower panel shows the photon index variation during the same period. {\it Bottom panels:} The photon index vs. the flux above $208.6$ MeV for F1 (left) and F2(right).
   }%
%%    \vspace{-1.6em}
    \label{fgindex}
\end{figure*}

 The \gray light curve is calculated using the unbinned likelihood analysis method implemented in the {\it gtlike} tool. $(0.1-500)$ GeV photons are used in the analysis with the appropriate quality cuts applied in the previous section. Different model files are used to ensure that the possible contribution from sources within ROI are properly accounted for.
In the model file obtained from the whole-time analysis, the photon indices of all background sources are first fixed to the best guess values in order to reduce the uncertainties in the flux estimations, then those of the sources within ROI are considered as free parameters. In addition we analyzed the data accumulated during the one-month periods covering the major flares (01-30 October and 15 December 2016-15 January 2017). Then we fixed the spectral parameters of all background sources as in \citet{aaa16}. All approaches yielded essentially the same results. We used the latter model as the rising and decaying times of the first flare can be evaluated better. Given shorter periods are considered, the spectrum of NGC 1275 has been modeled using a power-law function with the index and normalization as free parameters. Since no variability is expected for the background diffuse emission, the normalization of both background components is also fixed to the values obtained for the whole time period.\\
Fig. \ref{fg2} (upper panel) shows the \gray light curve with three-day bin size. Despite the fact that the flux sometimes exceeded the averaged value presented in 3FGL $(\approx 2.26\times10^{-7}\:{\rm photons \: cm^{-2}\:s^{-1}})$, pronounced flaring activities were detected in October 2015 (hereafter Flare 1 [F1]) and in December 2016/January 2017 (hereafter Flare 2 [F2]). Starting from 22 October 2015 the daily averaged flux of NGC 1275 was above $10^{-6}\:{\rm photons \: cm^{-2}\:s^{-1}}$ and remained high for 5 days with a daily averaged maximum of $(1.48\pm0.20)\times10^{-6}\:{\rm photons \: cm^{-2}\:s^{-1}}$ observed on 24 October 2015. Another substantial increase in the \gray flux was observed on December 31, 2016 when the flux increased from about $(4\sim5)\times10^{-7}\:{\rm photons \: cm^{-2}\:s^{-1}}$ to $(2.21\pm0.26)\times10^{-6}\:{\rm photons \: cm^{-2}\:s^{-1}}$ within a day with a detection significance of $\sim21.5\sigma$.\\
The photon statistics allowed us to study these flares with denser time sampling (sub-day) for the first time. The shortest bin sizes have been chosen to ensure that {\it i)} the flare rise and decay periods are well constrained and {\it ii)} the detection significance for each bin exceeds the $\sim5\sigma$ limit. The statistics allowed to use bins with 8-hour intervals for F1 and 3-hour bins for F2. For example, from MJD 57317 to MJD 57322, the detection significance varied between $5.1\sigma$ and $13.1\sigma$, and from MJD 557753 to MJD 57754 it was between $5.3\sigma$ and $10.4\sigma$. The corresponding light curves are shown in the middle panels of Fig. \ref{fg2}.  In order to check if the likelihood fit has converged in each time bin, the plot of Npred/$\sqrt{\rm Npred}$ vs Flux/$\Delta {\rm Flux}$ is shown in the lower panels of Fig. \ref{fg2} for 8-hour (left) and 3-hour (right) bins. We verified that the fit has converged in the surrounding bins as well. As one can see, it seems there is a linear correlation without any declination, so the errors are an accurate representation of the observation.\\
Further, the \gray flux and photon index variations are investigated using a light curve generated by an adaptive binning method \citep{Lott2012}. In this method, the time bin widths are flexible and chosen to produce bins with a constant flux uncertainty. This method allows detailed investigation of the flaring periods, since at times of a high flux, the time bins are narrower than during lower flux levels, therefore rapid changes of the flux can be found. In order to reach the necessary relative flux uncertainty, the integral fluxes are computed above the optimal energy \citep{Lott2012} which in this case is $E_0$ = $208.6$ MeV.  Also, in order to improve the accuracy of the method, the flux of bright sources which lie close to NGC 1275 have been taken into account. This is done by providing the parameters of confusing sources during the adaptive binning light curve calculations. The light curve calculated assuming a constant 15\% uncertainty is shown in Fig. \ref{fgindex} (upper panels) for the period covering the large flares.
\section{Spectral analysis}\label{sec3}
\subsection{\fermi data}
The changes in the \gray photon index are investigated by analyzing the data from the following four periods:
\begin{itemize}
\item[{\it i)}] overlapping with the observation of Swift on 30 December 2016. Even if the Swift observations lasted $\sim960$ seconds, in order to increase the statistics, the \gray spectrum has been extracted for the period MJD 57752.75-57753.25 where the source has a comparable flux as revealed from the light curve with a 6-hour binning (Fig. \ref{fg2}, middle right panel).
\item[{\it ii)}] MJD 57754.00-57755.75, when the flux is relatively constant and it coincides with the observations of Swift on 01 January 2017.
\item[{\it iii)}] at the peak of F2 (MJD 57753.81), using the data accumulated for 3 hours.
\item[{\it iv)}] MJD 57442.32-57444.45, which corresponds to the quiet (steady) state in the X-ray and \gray bands.
\end{itemize}
The \gray spectrum of NGC 1275 has been modeled using a power-law function ($dN/dE\sim N_{0}\:E^{-\Gamma}$) where the normalization ($N_{\rm 0}$) and power-law index ($\Gamma$) are considered as free parameters. The best matches between the spectral models and events are obtained with an unbinned likelihood analysis implemented in {\it gtlike}. The spectral fitting results are summarized in Table \ref{tab:results}. After analyzing the data for each considered period, the SEDs are obtained by freezing the NGC 1275 photon index in the model file and separately running {\it gtlike} for smaller energy bins of equal width in log scale. The SED for each period is shown in Fig. \ref{spectra}. Although some features can be noticed, it is hard to make any conclusion because of large uncertainties in the estimated parameters.\\
We separately analyze the \fermi data to determine the energy of the highest energy photon detected from NGC 1275 using {\it gtsrcprob} tool and the model file obtained from the likelihood fitting. All spectral parameters of the sources within ROI are first fixed to the best fitting values obtained in the whole time analysis and then are left free. Both yielded identical results. In this case, additional care must be taken since IC 310, which is known to be a strong emitter in the VHE \gray band \citep{alek14a}, is only at a distance of $0.623^{\circ}$. So both sources are considered to estimate the probability whether the photon belongs to NGC 1275 or to IC 310. The highest-energy photons detected during the four periods mentioned above are presented in Table \ref{tab:results}.
\subsection{Swift UVOT/XRT data}
During F2, Swift \citep{Gehrels2004} had observed NGC 1275 three times (see Table \ref{tab:results}). Unfortunately, there were no observations overlapping with F1. In addition to these observations, the Swift data of February 25, 2016, corresponding to a relatively stable state in the X-ray band have been analyzed. The XRT data were analyzed with the {\it XRTDAS} software package (v.3.3.0) distributed by HEASARC along with the HEASoft package (v.6.21). The source region was defined as a circle with a radius of 10.6 pixels ($25''$) at the center of the source, while the background region as an annulus centered at the source with its inner and outer radii being 20 ($47''$) and 30 pixels ($71''$), respectively. Such selections allowed to minimize the possible contribution from the cluster emission. For PC mode observations (Obsid 87312001, 87311001), the count rate was above 0.5 count/s, being affected by the piling-up in the inner part of the PSF. This effect was removed, excluding the events within a 4 pixel radius circle centered on the source position. All spectra were re-binned to have at least 20 counts per bin, ignoring the channels with energy below 0.3 keV, and fitted using {\it Xspec} v12.9.1a. The results of the fit are presented in Table \ref{tab:results}.\\
In the analysis of Swift-UVOT data, the source counts were extracted from an aperture of $5.0''$ radius around the source. The background counts were taken from the neighboring circular source-free region with a radius of $20''$. The magnitudes were computed using the {\it uvotsource} tool (HEASOFT v6.21) corrected for extinction according to \citet{roming}, using $E(B-V)=0.14$ from \citet{schlafly} and zero points from \citet{breeveld} converted to fluxes, following \citet{poole}. The corresponding spectra are shown in Fig. \ref{spectra}.
\section{RESULTS and DISCUSSIONS}\label{sec4}
\begin{centering}
\begin{figure}
    \includegraphics[width= 0.5 \textwidth]{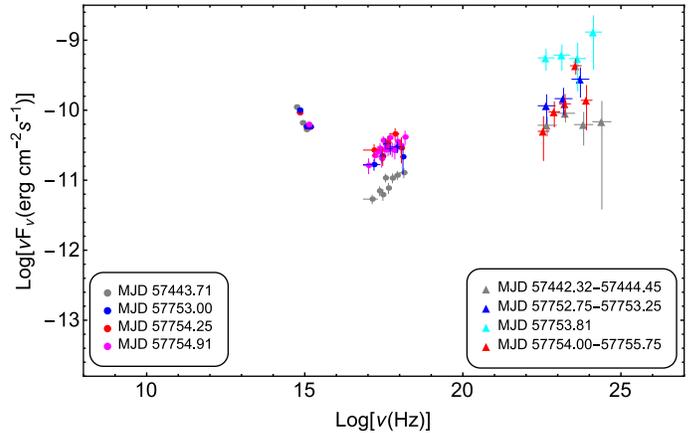}
    \caption{The multiwavelength SED for the periods presented in Table \ref{tab:results}.}%
    \label{spectra}
\end{figure}
\end{centering}

The \gray emission from one of the brightest radio galaxies in the MeV/GeV band- NGC 1275- has been investigated using the \fermi data accumulated during the last $\sim8.7$ years. The \gray light curve appears to be quite a complex one, with many peaks and flaring periods. The highest fluxes were detected in October 2015 and December 2016/January 2017 when the daily averaged peak \gray fluxes $\simeq(1.48-2.21)\times10^{-6}\:{\rm photons \: cm^{-2}\:s^{-1}}$ integrated above $100$ MeV were detected. It reached its maximum of $(3.48\pm0.87)\times10^{-6}\:{\rm photon\:cm^{-2}\:s^{-1}}$ on December 31, 2016, within 3 hours, which is the highest \gray flux observed from NGC 1275 so far; it exceeds the averaged flux by a factor of $\sim15.4$. The apparent isotropic \gray luminosity at the peak of the flare, $L_{\gamma}\simeq3.84\times10^{45}\:{\rm erg\:s^{-1}}$ (using $d_{\rm L}=75.6$ Mpc), exceeds the averaged \gray luminosity of other radio galaxies detected by \fermi (usually $\leq10^{45}\:{\rm erg\:s^{-1}}$ \citep{abdomis}); it is more comparable with the luminosity of BL Lac blazars. This is quite impressive, considering the large Doppler boosting factors of blazars ($\delta\geq10$) as compared with the value of $\delta\sim(2-4)$ usually used for the radio galaxies. Yet, at $\delta=4$ the total power emitted in the \gray band in the proper frame of the jet would be $L_{{\rm em,}\gamma}\simeq L_{\gamma}/2\:\delta^2\simeq 1.2\times10^{44}\:{\rm erg\:s^{-1}}$. It is of the same order as the kinetic energy of the NGC 1275 jet ($L_{\rm jet}\simeq (0.6-4.9)\times10^{44}\:{\rm erg\:s^{-1}}$) estimated from broadband SED modeling \citep{abdo2009}. This implies that during the discussed flaring period a substantial fraction of the total jet power, ($L_{{\rm em,}\gamma}/L_{\rm jet}\leq1$), is converted into $\gamma$-rays. These assumptions are in a strong dependence with $\delta$, which is highly unknown. But it seems that $\delta=4$ is already a limiting case, and larger decrease of $L_{{\rm em,}\gamma}$ is not expected.\\
The \gray spectrum estimated during the peak flux appeared as a nearly flat one (cyan data in Fig. \ref{spectra}), though the photon index estimation uncertainty is large ($\Gamma=1.93\pm0.19$). This is similar with the spectrum measured in a quiet state (Fig. \ref{spectra}, gray data), although with a significantly increased flux. The \gray photon index measured before and after the peak flux hints at spectral hardening (see Table \ref{tab:results} and Fig. \ref{spectra} blue and red data). However, large uncertainties in the photon index estimations do not allow us to make strong conclusions on the spectral hardening or softening. Although, as compared with the quiet state, it is clear that during the active states the \gray flux increases and the spectrum shifts to higher energies.\\
The broadband SED of NGC 1275 (Fig. \ref{spectra}) shows that during the bright \gray states, the X-ray flux also has increased. The analysis of the Swift XRT data detected during F2 results in an unabsorbed flux of $F_{0.3-10\:{\rm keV}}\approx(0.86-1.06)\times10^{-10}\:{\rm erg\:cm^{-2}\:s^{-1}}$ which nearly 3 times exceeds the averaged flux observed in February 2016. We note that the X-ray photon index measured during the quiet state is somewhat similar to the values measured by XMM-Newton \citep{churazov} and Swift BAT \citep{ajelo} while during the active states the X-ray photon index is steeper ($\sim1.7$). In the lower energy band, the UV flux from UVOT observations is relatively stable when comparing the quiescent and flaring states, albeit the data from all filters are not available to make definite conclusions.
\begin{deluxetable*}{lcccccc}
\tablecaption{Parameter values best explaining the flares. \label{fit_par}}
\tablehead{
\colhead{Flare period}  & \colhead{$t_r \pm err$} & \colhead{$t_d \pm err$} & \colhead{$t_{0}$} & $F_{\rm c}$ & $F_{\rm 0}$  \\
\colhead{} & \colhead{(hour) } & \colhead{(hour)} & \colhead{MJD} & \colhead{$\times 10^{-7}\mathrm{\rm photon\:cm}^{-2}s^{-1}$} & \colhead{$\times 10^{-7}\mathrm{\rm photon\:cm}^{-2}s^{-1}$}
}
%\colnumbers
\startdata
2015 October &  $32.49 \pm 7.20$ & $2.22\pm1.19$ & $57320.41\pm 0.19$  & $8.43\pm1.42$ & $23.92\pm3.08$  \\
2016 December/2017 January &  $8.03\pm1.24$ & $1.21\pm0.22$ & $57753.88\pm 0.04$  & $9.73\pm1.75$ & $41.96\pm4.82$  \\
\enddata
\end{deluxetable*}
\subsection{\gray photon index variation}
 The \gray photon index changes during $\sim8.7$ years of \fermi observations are investigated with the help of an adaptively binned light curve. In Fig. \ref{fgindex} (upper panels) the photon flux and index variation in time are shown for the time that covers only F1 and F2. In the course of $\sim8.7$ years, the hardest photon index of $\Gamma=1.62\pm 0.13$ was observed on MJD $55331.51$ for $\sim2.78$ days, while the softest index of $\Gamma=2.77\pm0.21$ was detected on MJD $56124.71$. The lowest and highest fluxes (above $208.6$ MeV) were $F_{\gamma}=(4.27\pm1.06)\times10^{-8}\:{\rm photon\:cm^{-2}\:s^{-1}}$ and $F_{\gamma}=(1.18\pm0.28)\times10^{-6}\:{\rm photon\:cm^{-2}\:s^{-1}}$, respectively. When the source is in active state, the data accumulated for a few hours is already enough to reach 15\% flux uncertainty, while in the quiet states, the data should be accumulated for several days. Interestingly, in the first $\sim 8.7$ years of \fermi operation, the highest-energy photon with $E_\gamma=241.2\:{\rm GeV}$ has been detected on MJD 57756.62 (after F2) within a circle of $0.071^{\circ}$ around the nucleus of NGC 1275 with the $3.36\sigma$ probability to be associated with it. Another events with $E_\gamma=221.5, 164.9, 125.6, 123.3$  and 109.2 GeV have been observed on MJD 55402.39, 56760.82, 56610.75, 56578.00 and 57694.65, respectively. We note that the PSF of \fermi at energies $>10\; {\rm GeV}$ is sufficient to distinguish the photons with high accuracy, so the highest energy photons are most likely coming from NGC 1275. It appeared that the \gray spectra for the periods when the highest energy photons were emitted, have mostly harder photon indexes (e.g., $\Gamma=1.74\pm0.14$ when $E_\gamma=241.2$ GeV photon was detected). Likewise, when photons with $E_\gamma=221.5, 164.9, 125.6, 123.3$ and 109.2 GeV were detected, the photon indexes were $\Gamma=1.81\pm0.15, 1.93\pm0.15, 1.79\pm0.13, 1.94\pm0.14$ and $1.86\pm0.15$, respectively. This hardening is probably associated with the emission from reaccelerated or fresh electrons, which produce also the observed highest energy photons.\\
The spectral changes observed in the photon-index-flux plane give us important information about the dynamics of the source and an insight into the particle acceleration and emission processes. The photon index $\Gamma$ as a function of the flux during F1 and F2 is shown in the bottom panels of Fig. \ref{fgindex}. A counter-clockwise loop is observed during F1, while during F2 the spectral index and flux changes follow a clockwise path. Such loops are expected to occur as a consequence of diffusive particle acceleration at strong shocks and cooling of the radiating particles. As discussed in \citet{krm98}, it is expected to have a counter-clockwise loop, if the variability, acceleration and cooling timescales are similar, implying that during the flare, the spectral slope is controlled by the acceleration rather than by the cooling processes. Consequently, the occurrence of a flare propagates from lower to higher energies, so the lower energy photons lead the higher energy ones. Instead, if the spectral slope is controlled by synchrotron cooling or any cooling process that is faster at higher energies, a clockwise loop will be seen. The counter-clockwise loop observed during F1 suggests that, most likely, this flaring event is due to the acceleration of the lower-energy electrons. Note that such `harder-when-brighter' behavior was already observed during the previous flares of NGC 1275 \citep{kataoka2010, brown}. The clockwise loop observed during F2 indicates that during this flare the flux started to increase at low energies (HE radiating particles cool down and radiate at lower and lower energies) and then propagate to HE. This shows that HE electrons are playing a key role during F2, which also produce the highest energy photons from NGC 1275 observed around F2.\\
The interpretation of the mechanism responsible for spectral evolution can be more complicated than it was discussed above. It has been already shown that, depending on the change of the total injected energy, the dominance of synchrotron and Compton components can also vary, so that the trajectory in the photon index-flux plane evolves clockwise or counterclockwise, depending on the total energy and the observed energy bands \citep{botcher,li}. Thus, the observed spectral evolution is quite sensitive to various parameters in the model and it is hard to draw any firm conclusions. The discussions above are of first order approximation and are generally meant to understand the dynamics of the system.
\subsection{Minimum flux variability period:} During F1 and F2, the flare time profiles are investigated by fitting them (Fig. \ref{fg2} middle panels blue data) with double exponential functions in the following form \citep{abdovar}:
\begin{equation}
F(t)= F_{\rm c}+F_0\times\left(e^{\frac{t-t_{\rm 0}}{t_{\rm r}}}+e^{\frac{t_{\rm 0}-t}{t_{\rm d}}}\right)^{-1}
\label{func}
\end{equation}
where $t_{0}$ is the time of the maximum intensity of the flare ($F_0$) and $F_c$ is the constant level present in the flare. $t_{\rm r}$ and $t_{\rm d}$ are the rise and decay time constants, respectively. The fitting parameters are summarized in Table \ref{fit_par} and the corresponding fit is shown in Fig. \ref{fg2} middle panels (red dashed line). The time profiles show asymmetric structures in both flares, showing a slow rise and a fast decay trend. The time peak of the flares calculated by $t_{\rm p}=t_{0}+t_{\rm r}\:t_{\rm d}/(t_{\rm r}+t_{\rm d})ln(t_{\rm d}/t_{\rm r})$ is MJD $57320.18$ for F1 with the maximum intensity of $(2.39\pm0.31)\times 10^{-6}\;\mathrm{photon\:cm}^{-2}s^{-1}$. The rise time is $32.49 \pm 7.20$ hours with a sudden drop within $2.22\pm1.19$ hours. The parameters of F2 are better estimated and are characterized with a shorter rise time, when within $8.03\pm1.24$ hours the flux reaches its maximum of $(4.20\pm0.48)\times10^{-6}\:\mathrm{\rm photon\:cm}^{-2}s^{-1}$ on MJD $57753.79$ and drops nearly 4 times in $\sim6$ hours. The minimal e-folding time is $t_{\rm d}=1.21\pm0.22$ hours, using the decay time scale of F2, and it is the most rapid \gray variability observed for NGC 1275. We note that even if the rise time of F2 is used, the flux e-folding time of about $8.03\pm1.24$ hours will still be shorter than any previously reported value.\\
The obtained shortest flux e-folding time, $t_{\rm d}=1.21\pm0.22$ hours, is unusual for radio galaxies and has never been observed for any other radio galaxy so far. It is more similar to the rapid \gray variability detected from several bright blazars \citep{brown13,Foschini11,Foschini13, Saito13, nalewajko, hnm15, Rani13, aaa16}. \citet{brown13} was the first to point out that during the \gray flares of PKS 1510-089 the flux doubling time-scale was as short as $1.3\pm0.12$ hours which was the shortest variability time-scales measured at MeV/GeV energies at that time. It is interesting that such rapid \gray variability is mostly observed from flat-spectrum radio quasars. The asymmetric profile of NGC 1275 flares can be explained if assumed that the accelerated particles (e.g., by shock acceleration) quickly cool down due to the increase of the magnetic field (assuming the electrons dominantly lose energy by synchrotron cooling). In order to interpret the fast decay ($t_{\rm dec}=1.21\pm0.22$ hours) as cooling of relativistic electrons ($t_{\rm decay}=t_{\rm cooling}/\delta$; $t_{\rm cooling}=6\:\pi \:m_e^2\:c^{3}/\sigma_{\rm T} B^2\:E_{e}$) with $E_e=100\:{\rm GeV}$, the magnetic field should be $B\approx478\;{\rm mG}\:(\delta/4)^{-1/2}(t_{\rm dec}/1.2\:h)^{-1/2}\:(E_{e}/100 GeV)^{-1/2}$ (where we assumed a moderate Doppler boosting factor of $\delta=4$), which is not far from the typical values usually used in the modeling of emission from radio galaxies \citep{abdom87,abdocena}. Even if the magnetic field is $10-100$ times lower than this value, the shock acceleration time scales ($t_{\rm acc}\approx 6\:r_g\:c/v_{s}^2$ \citep{Rieger07}) would be more than enough to accelerate the electrons $>\:100\:{\rm GeV}$ within the observed rise time scale ($8.03\pm1.24$ hours).
\subsection{The origin of emission}
The observed short time scale variability of $1.21\pm0.22$ hours allows to constrain the characteristic size of the emitting region radius to $R_{\gamma}\leq\:\delta\times c\times \tau\approx5.22\times10^{14}\:(\delta/4)\:{\rm cm}$. {If the entire jet width is responsible for the emission,} assuming the jet half-opening angle $\theta_{\rm j}\simeq 0.1^\circ$, the location of the emitting region along the jet will be $r\simeq R_{\gamma}/\theta_j \simeq0.1\:(\delta/4)\:(\theta_{j}/0.1^\circ)^{-1}\:{\rm pc}$. This strongly suggests that the observed emission is most likely produced in the subparsec-scale jet. In principle the jet can be much more extended and the emission is produced in a region smaller than the width of the jet. For example, multiple regions moving in a wider jet having different beaming factors  can be an alternative possibility \citep{lenain}. In this model, the emission is expected to take place in a broadened jet formation zone close to the central supermassive black hole, where even for a large jet angle, a few emission zones can move directly toward the observer and Doppler boost the emission. Here the emission region is very close to central source, again implying that the innermost jet (subparsec-scale) is responsible for the emission.\\
The SED presented in Fig. \ref{spectra} as well as that in \citet{aleksic2014}, hint at a double-peaked SED similar to those of other GeV/TeV-emitting radio galaxies \citep{abdom87,abdocena} and blazars. This similarity allowed to model the SED of NGC 1275 within the one-zone synchrotron SSC scenario \citep{aleksic2014}. However, it failed to reproduce the required large separation of the two peaks (gray data in Fig. \ref{spectra}) with small Doppler factors ($\delta=2-4$) typical for radio galaxies. With the new data, the situation even worsened: even if the data are not enough to exactly identify the location of the peaks, clearly, the first peak is at $\sim(10^{14}-10^{15}) {\rm Hz}$ (unchanged) while the rising shape of the MeV/GeV spectrum indicates the second peak shifted to higher frequencies. Such large separation of the two SED peaks unavoidably requires a higher Doppler factor than that used previously.  Moreover, if one-zone SSC emission dominates, usually it is expected to have correlated changes in the X-ray/\gray band, which are not observed here. One can avoid these difficulties by assuming that HE emission is produced in a local substructure of the jet, which is characterized by a higher boosting factor and/or smaller inclination angle. For example, the mini-jets generated by local reconnection outflows in a global jet (`jets in a jet' model \citep{gub09}) can have extra Lorentz boosting and the emission can be produced around these local reconnection regions. This successfully explains the fast TeV variability of M87 \citep{gub10} so that it can be naturally considered also in this case. In addition, two-zone SSC models, when different regions are responsible for low and high-energy emissions, can be an alternative. In more complex-structured jet models the seed photons for IC scattering can be of external origin (the emission region is the layer and external photons are from the spine, or vice versa \citep{tavecchioa14}) the energy of which is higher than that of synchrotron photons resulting in the shift of the emission peak to higher energies. However, these models involve additional parameters, which cannot be constrained with the current data set and additional observations in the radio/optical and VHE \gray bands are required.
\section{SUMMARY}\label{sec5}
We report on the results of $\sim8.7$ years' \gray observations of NGC 1275 radio galaxy. The source displayed  prominent flaring activities in October 2015 and December 2016/January 2017 with the 3-hour peak flux above 100 MeV of $(3.48\pm0.87)\times10^{-6}\:{\rm photon\:cm^{-2}\:s^{-1}}$ observed on 2016 December 31 corresponding to an apparent isotropic \gray luminosity of $L_{\gamma}\simeq3.84\times10^{45}\:{\rm erg\:s^{-1}}$. This luminosity is more typical for BL Lac blazars and corresponds to a large fraction of the kinetic energy of the NGC 1275 jet, implying that the \gray production efficiency is very high.\\
During the major flares, the photon statistics allowed us to investigate the flare properties with as short as 3-hour intervals for the first time. This allowed to find very rapid variability with the flux e-folding time as short as $1.21\pm0.22$ hours, which is very unusual for radio galaxies. The \gray photon index of the source was evolving during the flaring periods, showing counter clockwise and clockwise loops in the photon-index-flux plane during the flares in October 2015 and December 2016/January 2017, respectively. Also, some of the highest energy \gray photons observed from the source during $\sim8.7$ years arrived around the same active periods. Perhaps this rapid \gray flare was associated with effective particle acceleration that led to emission of these photons.\\
The observed hour-scale variability suggests that the emission is produced in a very compact emission region with $R_\gamma\leq5.22\times10^{14}\:(\delta/4)\:{\rm cm}$, and perhaps it is produced in a sub-parsec scale jet. During the \gray activity, the HE component not only increased but also shifted to higher energies. Considering this shift and the large \gray luminosity, it makes it very challenging to explain the observed X-ray and \gray data in the standard synchrotron/SSC models. Additional assumptions on the jet structure/emission processes are to be made.
\section*{acknowledgements}
This work was supported by the RA MES State Committee of Science, in the frames of the research project No 15T-1C375. We thank the anonymous referee for constructive comments that significantly improved the paper.
\bibliographystyle{apj}

\end{document}